\documentclass[a4paper,twoside]{article}
\usepackage{deluxetable}
\newcommand{\affil}[1]{$^{\rm #1}$}
\usepackage[authoryear]{natbib}
\bibpunct{ (}{)}{;}{a}{}{,}
\usepackage{graphicx}
\usepackage{epsfig}
\usepackage{epstopdf}
\usepackage{gensymb}
\date{} 
\newcommand{\kms}{\mbox{km\,s$^{-1}$}}
\def\kms {\ifmmode{{\rm ~km~s}^{-1}}\else{~km~s$^{-1}$}\fi}
\def\lsun {\ifmmode{{\rm ~L}_\odot}\else{~L$_\odot$}\fi}

\newbox\grsign \setbox\grsign=\hbox{$>$} \newdimen\grdimen \grdimen=\ht\grsign
\newbox\simlessbox \newbox\simgreatbox
\setbox\simgreatbox=\hbox{\raise.5ex\hbox{$>$}\llap
 {\lower.5ex\hbox{$\sim$}}}\ht1=\grdimen\dp1=0pt
\setbox\simlessbox=\hbox{\raise.5ex\hbox{$<$}\llap
 {\lower.5ex\hbox{$\sim$}}}\ht2=\grdimen\dp2=0pt

\def \etal {\rm ~{\it \etal},~}

\title{\large\bf\flushleft {Australian Aboriginal Astronomy: Overview}}
\author{\parbox{\textwidth}{\flushleft
\vspace{-0.5cm}
{\it
Ray P.\ Norris\affil{1,2},
Duane W. Hamacher\affil{3}
}
 \\
\vspace{0.4cm}
{\small \affil{1}\,Department of Indigenous Studies, Macquarie University, NSW, 2109, Australia}\\ 
{\small \affil{2}\,CSIRO Astronomy \& Space Science, PO Box 76, Epping, NSW, 1710, Australia\\ 
email: {\tt RayPNorris@gmail.com}}\\
{\small \affil{3}\,Nura Gili Centre for Indigenous Programs, University of New South Wales, Sydney, NSW, 2052, Australia}\\ 
}}
\begin{document}
\twocolumn[
\begin{changemargin}{.8cm}{.5cm}
\begin{minipage}{.9\textwidth}
\vspace{-1cm}
\maketitle
%
%

{\bf Abstract:} 
The traditional cultures of Aboriginal Australians include a significant astronomical component, perpetuated through oral tradition, ceremony, and art. This astronomical component includes a deep understanding of the motion of objects in the sky, and this knowledge was used for practical purposes such as constructing calendars. There is  also evidence that traditional Aboriginal Australians made careful records and measurements of cyclical phenomena, paid careful attention to unexpected phenomena such as eclipses and meteorite impacts, and could determine the cardinal points to an accuracy of a few degrees.


\medskip
\medskip
\end{minipage}
\end{changemargin}
]
%

\section{Introduction}
\label{intro}
Humans first came to Australia  \index{Australia} \index{Aboriginal}
at least 40,000 years ago \citep{OConnell} and enjoyed a continuous, unbroken, culture until the arrival of the British in 1788, making Aboriginal Australians among the oldest continuous cultures in the world \citep{mcniven}. At the time of British occupation, there were about 300 distinct Aboriginal language groups with nearly 750 dialects \citep{Walsh}. Each had its own language, stories, and beliefs, although most were centred on the idea that the world was created in the ``Dreaming'' \index{Dreaming} by ancestral spirits, whose presence can still be seen both on the land and in the sky.  According to traditional songs and stories, these spirits taught humans how to live, and laid the basis for a complex system of law and morality. 

Most Aboriginal Australians were nomadic hun\-ter/\-gatherers, each clan moving in an annual cycle to seasonal camps and hunting-places within the land that they owned, to take advantage of seasonal food sources. They practised careful land management to increase the food yield of their land \citep{Gammage}, including ``firestick--farming'' in which the land was burnt in a patchwork pattern to encourage young growth and mitigate the effect of unplanned fires. Some groups built stone traps for fish farming, planted crops such as yams, or built stone dwellings \citep{clark94}.

Most of the Aboriginal cultures were severely damaged by the arrival of Europeans, particularly in southeastern Australia. However, some groups in the north and centre of Australia still retain almost all of their pre-contact language and culture, and still conduct initiation ceremonies where knowledge is passed from one generation to the next. It is from these groups, particularly the Yolngu and Wardaman people, that we have obtained our most detailed information. As some of this material is sacred, and inaccessible to non-Aboriginal people, this article refers only to material which has been designated as ``public'' by traditional Aboriginal owners.

The songs, stories, art, and ceremonies of many traditional Aboriginal cultures refer to the Sun, Moon, planets, stars, the Milky Way, and the dark clouds within it \citep{stanbridge,mountford, haynes, johnson, harney, emudreaming, norris09, norris11,peru}. For example, Aboriginal ``constellations'' include the Southern Cross, which in different Aboriginal cultures might represent an emu footprint, a stingray, or a possum in a  tree. On the other hand, in many different Aboriginal cultures, Orion symbolises a young man or group of young men, chasing the Pleiades (Seven Sisters). Another well-known ``constellation'', the emu \citep{harney, massola}, consists not of stars but of the dark clouds within the Milky Way, and is important in many different cultures across Australia.

\begin{figure}[hbt]
\includegraphics[width=8cm]{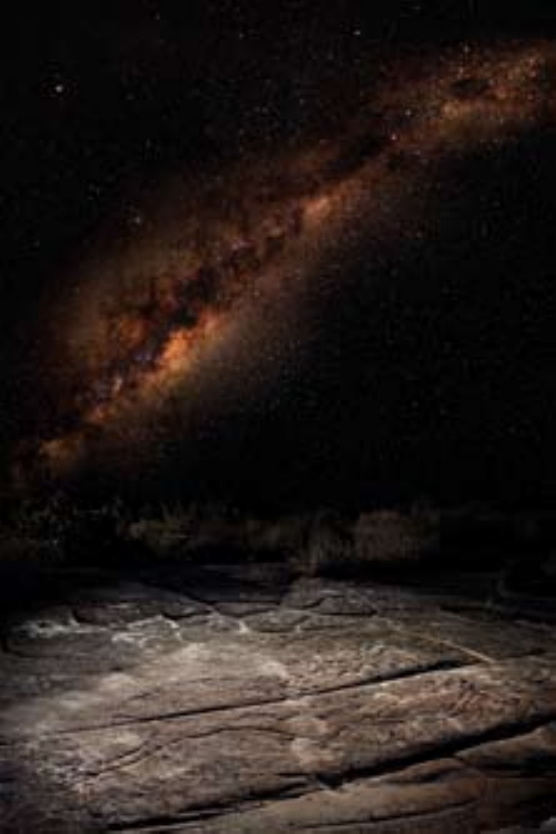}\vspace{2mm}
\caption{An image of the Emu in the Sky, marked by dark clouds in the Milky Way, directly above an engraving  which may depict it (Hugh Cairns, priv. comm). Photograph courtesy Barnaby Norris} 
\label{fig1}
\end{figure}
 

Of particular interest to ethnoastronomy is the deeper understanding of the sky, such as explanations of tides, eclipses, the motion of the Sun and Moon, and the ability to predict the rising and setting places of celestial bodies. The sky also had practical applications for navigation and time keeping \citep{harney, Clarke09}, discussed in greater detail by \cite{Clarke13}.


\section{Aboriginal Astronomy}
\subsection{The Sun, Moon, and Planets}
\index{Sun} \index{Moon} 
Many of the stories show that traditional Aboriginal people sought to understand the motion of celestial bodies, and to place them in a self-consistent framework that described the natural world. For example, the motion of the Sun is described by Yolngu people as being caused by the Sun-lady, Walu, who each morning lights a stringy-bark tree and carries it across the sky to her camp in the West \citep{Wells64}. The Moon-man, Ngalindi, is attacked with axes by his wives, slicing pieces off him and causing the lunar phases. 

\index{eclipse}
Solar eclipses were widely viewed as a bad omen, but there is also good evidence that at least three Aboriginal groups (the Euahlayi, Yolngu, and Warl\-piri people, each from a  different state of Australia) recognised that an eclipse was caused by a conjunction between the Sun and Moon \citep{eclipse}. In all three cases they interpreted it as a mating between the Sun-woman and the Moon-man, with one account (the Warlpiri) adding that  the Sun-woman was sent away by the sky-spirits because of her attempted seduction of the Moon-man. Many other Aboriginal groups recognised that an eclipse was caused by something covering the Sun, but thought that the ``something'' was a hood or cloak. A total solar eclipse is seen in any one location only once every three or four generations, and so these explanations imply a remarkable continuity of learning. 

Lunar eclipses were also widely interpreted as something covering the Moon, but in only one case was it attributed to the relative positions of the Sun and Moon\citep{eclipse}. In other cases, the Moon-man is thought to be covering his face, or the Moon is covered by the shadow of  a Man who is walking in the Milky Way. The red colour of the Moon during a lunar eclipses was widely attributed to the Moon-man having blood on his face.

\index{tides} 
The height of the tides varies with the phase of the Moon, with the highest tide (``spring tide'') occurring at the new moon, an observation that Galileo failed to notice. The Yolngu did notice this connection and devised an explanation for the phenomenon. Their explanation, based on the Moon filling and emptying as it passed through the ocean at the horizon \citep{Berndt}, is rather different from modern science but is nevertheless a good example of an evidence-based approach to understanding the world in an appropriate cultural context. 

\index{Venus}
Similarly, the planets were widely seen to move differently from the stars. For example, the Yolngu noted that Venus was always low in the sky, close to the rising or setting sun. They explained this by suggesting that Venus as a morning star was attached by a rope to the mythical island of Baralku in the east \citep{emudreaming, Allen}, preventing her from rising high in the sky. Similarly, Venus as an evening star was held down by a rope connected to the ``spirits in the West'' \citep{Berndt}. It's possible that the zodiacal light, which is easily visible from Arnhem Land, was seen as supportive evidence for this rope.


\subsection{Orientation and Prediction}
\index{Cardinal Directions}
The concept of cardinal directions is important to several Aboriginal groups, most notably the Warl\-piri people, much of whose cultural lore is based on cardinal directions \citep{Wanta}, largely determined by  the rising and setting sun. In some Aboriginal cultures, dead people are buried facing east \citep[e.g.][]{Mathews04} and initiation sites are often oriented north-south\citep{fuller}. \citet{rows} have shown that a sample of linear stone arrangements  are oriented north-south with an accuracy of a few degrees. 

Assuming that a magnetic compass was not available to the builders of these Boras and stone arrangements, three techniques could in principle be used to determine cardinal directions. One is the use of so-called ``magnetic" termite mounds, whose elongated shapes are aligned north-south with an accuracy of about ten degrees, to minimise solar heating of the mound (Grigg \& Underwood 1977). However, these are found only in the Northern Territory, far from the Boras and stone arrangements discussed here, and so a different technique must have been used. The remaining two techniques both use astronomical observations:

\begin{itemize}

\item From a given viewing position, the position of the setting sun may be marked each day with a stone or stick. The ends of the resulting line of markers will indicate the position of the Sun at the solstices, and the mid- point between these ends indicates due west. Ruggles (1997) has shown that the accuracy of this technique is limited by variations in the height of the horizon, but as the accuracy being cited here is of the order of a few degrees, this is unlikely to limit these measurements. We note that at Wurdi Young, discussed below, the solstitial positions are indeed marked, with due west being marked between them.

\item Similarly, the position of a circumpolar star, such as those within the Southern Cross, may be marked by placing a stick or stone vertically below the star at various times through the year. The midpoint of the resulting line will indicate due South.

\end{itemize}

\index{Wurdi Youang}
The best example of astronomical alignments is the Wurdi Youang stone ring \citep{morieson03, Norris12}, which has several east-west indicators, accurate to a few degrees, and outliers and relatively straight sections of the ring that indicate the setting position of the midwinter and midsummer sun. A Monte Carlo analysis indicates that these alignments are unlikely to be due to chance. This is the only Aboriginal site known to indicate significant astronomical positions on the horizon other than the cardinal points, and, unless it is a statistical freak or a hoax, suggests that other such sites may be discovered in the future.

\begin{figure}[hbt]
\includegraphics[width=8cm]{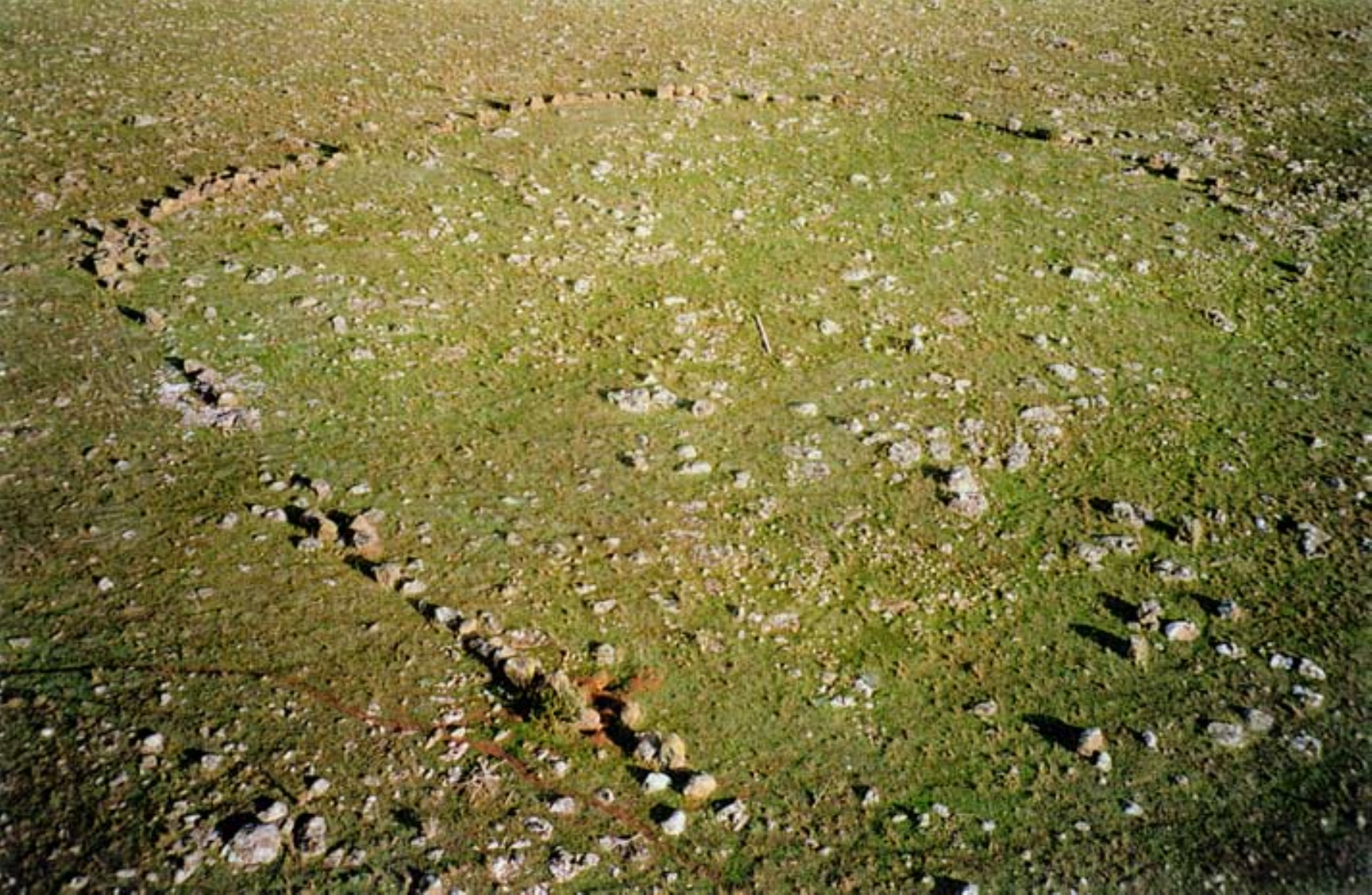}
\caption{Aerial view of the Wurdi Youang site, reproduced with permission from \cite{marshall}, looking west.} 
\label{aerial}
\end{figure}

\begin{figure}[hbt]
\includegraphics[width=8cm]{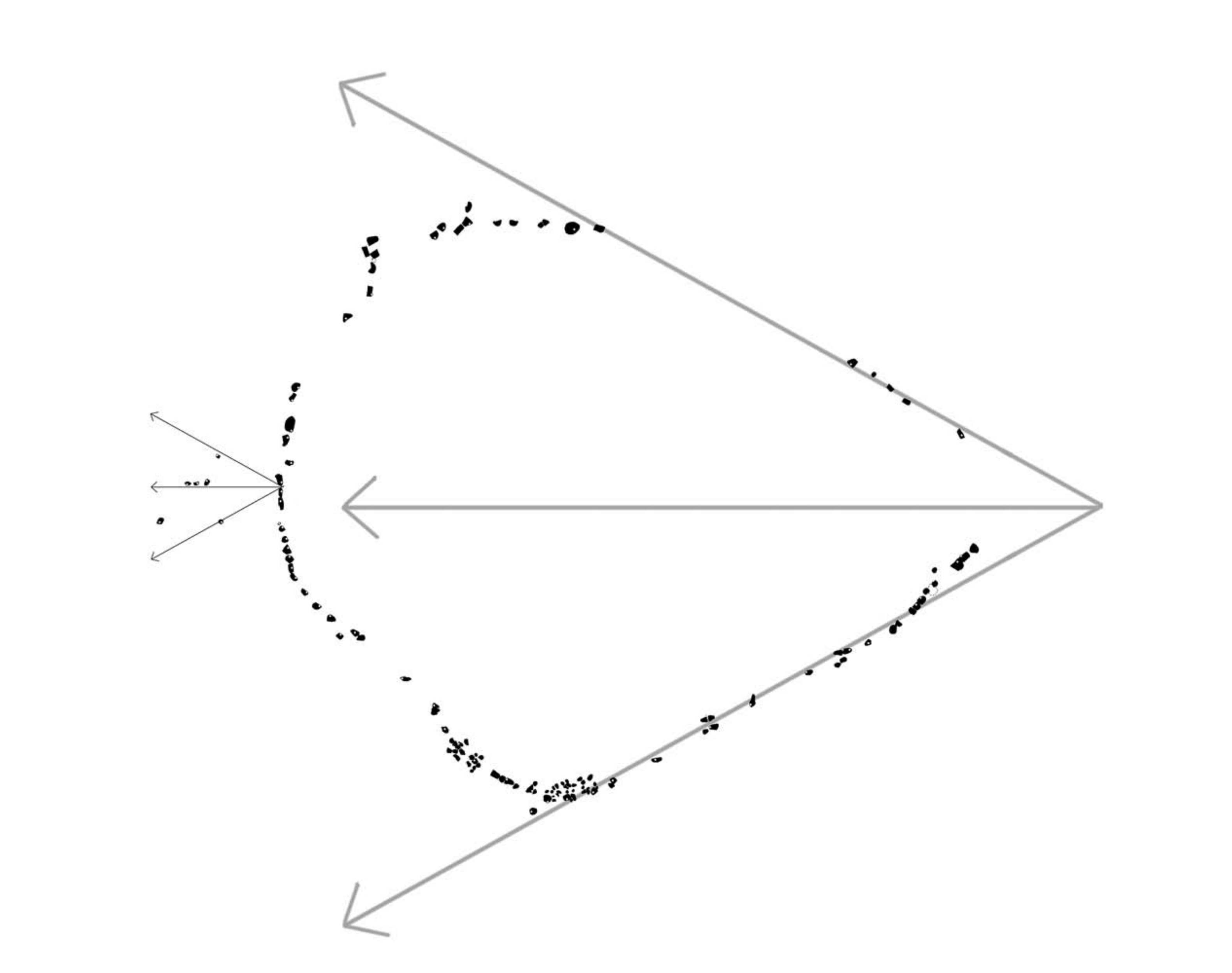}
\vspace{2mm}
\caption{The directions to the equinoxes and solstices superimposed on a plan of Wurdi Youang, taken from \cite{Norris12}.} 
\label{alignments}
\end{figure}

\subsection{Archaeoastronomical significance}
The evidence  shows unequivocally that Aboriginal Australians had a deep knowledge of the sky, and were aware of many celestial phenomena. There is also evidence that their interest in the sky went beyond this, to the extent of trying to understand the mechanisms behind these phenomena, and how they fitted into a self-consistent world view. 

These studies are hindered by the sparsity of data. Other than comets \citep{comets}, only one example has so far been found of a datable transient phenomenon that appears in a traditional Aboriginal oral account \citep[e.g. the Great Eruption of $\eta$ Carina: ][]{ecarina}, and attempts to link stories of stones from the sky with known meteorite events have so far been unsuccessful \citep{impacts}.
\index{meteor}

There is also a danger that we are trying to interpret Aboriginal culture through the very different eyes of Western culture, and may be imposing a Western world view on it. For example, it is sometimes asserted that  careful measurement is alien to Aboriginal culture, and until recently it was sometimes asserted \citep[e.g.][]{Blake} that no Aboriginal language had a word for a number greater than five. This latter assertion has no basis in fact, but appears to reflect a post-colonial prejudice combined with a lack of understanding of number systems. The best way to avoid such cultural bias, in either direction, is to rely only on evidence-based studies.

\subsection{Future directions}
Although the first evidence of Aboriginal astronomy was presented over 150 years ago \citep{stanbridge}, it is curious that only in the last few years has there been a concerted scholarly attempt to study the breadth and richness of Aboriginal astronomy. It is likely that we have barely scraped the surface, and far more lies undocumented. The material that is being found is unsurprising for those who study the ethnoastronomy of other cultures, but is a powerful tool in overcoming some of the prejudices that linger on in white Australian society. Aboriginal Astronomy is proving valuable in building greater understanding of the depth and complexity of Aboriginal cultures.

Against this optimistic view, much of the evidence is from oral tradition, and all Aboriginal cultures are under significant threat. The elders who possess ancient knowledge grow old and pass away, taking their knowledge with them, and relatively few bright young people rise to continue the tradition. Even in cases where the tradition is strong, better education and exposure to media mean that traditional knowledge is not static but is being augmented by modern education and New-Age ideas. It can be difficult for a teenager to remember whether a piece of knowledge came from his/her grandfather or from YouTube.

\section{Acknowledgements}
We acknowledge and pay our respects to the traditional owners and elders, both past and present,  of the Aboriginal peoples of Australia. 

\clearpage
\setcounter{figure}{0}
\onecolumn
\bibliographystyle{aj}
\bibliography{paper_refs}

\clearpage

\end{document}